\documentclass[onecolumn,preprint,a4paper,nofootinbib]{revtex4}
\usepackage[mathscr]{euscript}
\usepackage{ulem}
\usepackage{caption}
\usepackage{amsmath}
\usepackage{multirow}
\usepackage{booktabs,tabularx}
\usepackage{threeparttable}
\usepackage{floatrow}
\usepackage[font=small,labelfont=bf,tableposition=top]{caption}
\usepackage{booktabs}
\usepackage{threeparttable}
\usepackage{graphicx}
\usepackage{subfigure}
\usepackage{epstopdf}
\usepackage[colorlinks=true,linkcolor=red]{hyperref}
\usepackage{bm}
\usepackage[figuresright]{rotating}
\allowdisplaybreaks[3]
\newcommand{\bea}{\begin{eqnarray}}
\newcommand{\eea}{\end{eqnarray}}
\newcommand{\beq}{\begin{equation}}
\newcommand{\eeq}{\end{equation}}
\newcommand{\nn}{\nonumber}
\def\/{\over}

\begin{document}

\title{Effects of acceleration on  interatomic interactions}
\author{Shijing Cheng$^{1}$, Wenting Zhou$^{2,}$\footnote{Corresponding author: zhouwenting@nbu.edu.cn}, and Hongwei Yu$^{3,}$\footnote{Corresponding author: hwyu@hunnu.edu.cn}}

\affiliation
{$^{1}$ School of Physics and Information Engineering, Shanxi Normal University, Taiyuan, 030031, China\\
$^{2}$ Institute of Fundamental Physics and Quantum Technology, Department of Physics, School of Physical Science and Technology, Ningbo University, Ningbo, Zhejiang 315211, China\\
$^{3}$ Department of Physics, Key Laboratory of Low Dimensional Quantum Structures and Quantum Control of Ministry of Education, and Hunan Research Center of the Basic Discipline for Quantum Effects and Quantum Technologies, Hunan Normal University, Changsha, Hunan 410081, China}

\begin{abstract}
The Unruh effect establishes a fundamental equivalence between acceleration and thermality by demonstrating that a uniformly accelerated ground-state detector undergoes excitation as if immersed in a thermal bath. In this paper, we investigate how acceleration influences  the interaction between two ground-state atoms that are synchronously and uniformly accelerated in vacuum with proper acceleration $a$ and coupled to a fluctuating electromagnetic field.
We find that the resulting  interaction potential  comprises  both diagonal components $(\delta E)^{jk}$ with $j=k$, which are present in both inertial and acceleration cases, and off-diagonal components $(\delta E)^{jk}$ with $j\neq k$, which arise exclusively due to acceleration and vanish in the inertial case. The dependence of each component on acceleration and interatomic separation $L$ generally differs. For small accelerations, the leading-order diagonal components of the van der Waals (vdW) and Casimir-Polder (CP) interaction potentials remain unchanged from their inertial counterparts, exhibiting the standard scaling behaviors $\sim L^{-6}$ and $\sim L^{-7}$, respectively. In contrast, the off-diagonal components scale as $\sim a^2L^{-4}$ in the vdW subregions and  $\sim a^2L^{-5}$ in the CP subregion. However, when the acceleration becomes sufficiently large, both diagonal and off-diagonal components of the vdW and CP interaction potentials are significantly modified, giving rise to entirely new interaction behaviors that deviate from those observed in the inertial case, whether in vacuum or thermal environments, indicating a breakdown of the acceleration-thermality equivalence established by the Unruh effect for single detectors.
\end{abstract}
\maketitle

\section{Introduction}

Vacuum fluctuations, a prominent feature distinguishing the quantum vacuum from its classical counterpart, are known to underlie various physical phenomena, for instance, the dispersion interaction between two electrically polarizable atoms~\cite{Craig98}. This ubiquitous interaction, which arises between virtually all types of matter, has been an enduring subject of research.
It plays a crucial role in our understanding of interactions in fundamental physics and has wide-ranging implications across disciplines including atomic and molecular physics, condensed matter, high-energy physics, chemistry, biology, and materials science~\cite{Bloch08,Klimchitskaya09,Israelachvili73,Roth96,Tkatchenko11,Defenu23,Woods16}.

It is now well understood that the dispersion interaction between two neutral atoms originates from their coupling to vacuum fluctuations: these fluctuations induce instantaneous dipole moments in each atom and establish correlations between them, resulting in an interaction potential that strongly depends on their separation. Notably, in free space, this interaction scales as $L^{-6}$ in the van der Waals (vdW) regime ($L\ll\lambda$, where $\lambda$ is the typical atomic transition wavelength), and as $L^{-7}$ in the Casimir-Polder (CP) regime ($L\gg\lambda$), due to retardation effects~\cite{Casimir-Polder48}. It has also been found that this interatomic interaction can be significantly modified in the presence of external influences that alter the vacuum field fluctuations, such as the presence of boundaries~\cite{Power82,Spagnolo06,Passante07,Peng23} or thermal environments~\cite{McLachlan63,Boyer96,Milonni96,Ninham98,Wennerstrom99,Goedecke99,Passante07,Cheng221,Cheng23}.

Non-inertial motion represents another important factor capable of modifying field fluctuations, thereby influencing matter-field interactions. A prominent example is the Fulling-Davies-Unruh (FDU) effect~\cite{Fulling73,Davies75,Unruh76}, which predicts that a uniformly accelerated detector in vacuum perceives the Minkowski vacuum as a thermal bath with Unruh temperature $T_U = a/(2\pi)$, where $a$ is the proper acceleration. This implies that accelerated observers perceive vacuum fluctuations as thermal in nature, suggesting that interatomic interactions may be similarly altered under acceleration.

Indeed, recent studies have begun to explore this idea. For example, the interaction between two ground-state atoms uniformly accelerated in vacuum, modeled as monopole-coupled to a scalar field, has been investigated~\cite{Marino14,Cheng22}. These works reveal that the interaction differs significantly from that between static atoms in vacuum. In particular, when the interatomic separation $L$ exceeds the characteristic acceleration scale $L_a = a^{-1}$, the interaction scales as $(aL^4)^{-1}$ in both the vdW and CP regimes. This is in stark contrast to the inertial case, where the interaction scales as $L^{-2}$ for $L \ll \lambda$ and as $L^{-3}$ for $L \gg \lambda$. Furthermore, the interaction in the accelerated case does not generally match that between static atoms in a thermal bath at temperature $T_U$, although it displays thermal-like behavior in some regimes.

However, in realistic scenarios, atoms couple to the electromagnetic field{s} via dipole interactions, rather than to scalar fields via monopole coupling. Thus, the fundamental nature of interatomic interactions under acceleration in realistic settings remains to be understood. A recent study~\cite{Zhou25} partially addressed this by examining two accelerated atoms with orthogonal polarizabilities, one aligned along the interatomic separation and the other perpendicular to it. The interaction in this setup exhibits several novel features: it emerges only under acceleration, vanishes in the inertial limit, does not appear in a thermal bath, displays unusual distance dependence, and even shows a duality in its acceleration dependence. Remarkably, the interaction can switch between attractive and repulsive as the separation varies.

While insightful, the cross-polarizable case considered in Ref.~\cite{Zhou25} is highly specific. In more general and realistic situations, atoms may possess dipole moments oriented in arbitrary directions. This raises the question: how does acceleration affect interatomic interactions in the most general case? In this work, we aim to address this question through a comprehensive investigation.
We consider two ground-state atoms that are uniformly and synchronously accelerated in vacuum, interacting with the quantum electromagnetic field. Using the fourth-order DDC formalism, we derive general expressions for the interaction potential, analyzing how both the atomic polarizability and the acceleration affect the interaction. Our results demonstrate that the interaction depends crucially on the directionality of atomic polarizability and on  acceleration-induced modifications to vacuum fluctuations. Notably, the equivalence between acceleration and thermality, established by the Unruh effect through the transition rates of single detectors, breaks down in the context of interatomic interactions.

This paper is organized as follows. In Section II, we introduce the setup and derive general expressions for the interaction potential using the fourth-order DDC formalism. Section III analyzes how uniform acceleration alters the interaction. Section IV summarizes our main findings. Throughout the paper, we use natural units with $\hbar = c = k_B = 1$.

\section{Interaction potential of two atoms uniformly accelerated in vacuum}

\begin{figure}[H]
\centering
\includegraphics[width=0.4\linewidth]{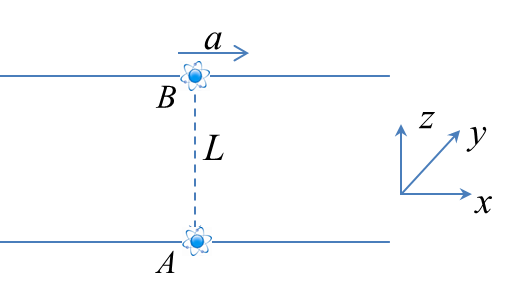}
\caption{Two atoms in synchronized uniform acceleration.}
\label{setup}
\end{figure}

As shown in Fig.~\ref{setup}, we consider two ground-state two-level atoms, labeled by $\xi = A, B$, undergoing synchronous and uniform acceleration in vacuum along the trajectories:
\bea
t_A(\tau)&=&\frac{1}{a}\sinh{(a\tau)}\;,\quad x_A(\tau)=\frac{1}{a}\cosh{(a\tau)}\;, \quad y_A=0\;, \quad z_A=0\;,\label{trajectoryA}\\
t_B(\tau)&=&\frac{1}{a}\sinh{(a\tau)}\;,\quad x_B(\tau)=\frac{1}{a}\cosh{(a\tau)}\;, \quad y_B=0\;, \quad z_B=L\;,\label{trajectoryB}
\eea
where $a$ is the proper acceleration, $\tau$ is the proper time, and $L$ is the constant interatomic separation.

In vacuum, the atoms are inevitably coupled to the fluctuating electromagnetic field. The total Hamiltonian of the combined ``atoms + field" system consists of three components:
\beq
H(\tau)=H_S(\tau)+H_F(\tau)+H_I(\tau)\;,
\eeq
where  $H_S(\tau)$ describes the atomic system, $H_F(\tau)$ the electromagnetic field, and $H_I(\tau)$ the interaction between the atoms and the field. The Hamiltonian of the atomic system is given by
\bea
H_S(\tau)&=&\sum_n\omega^A_n\sigma_{nn}^A(\tau)+\sum_n\omega^B_n\sigma_{nn}^B(\tau)\;,
\eea
where $\sigma_{nn}=|n\rangle\langle n|$ with $|n\rangle=|g\rangle$ and $|e\rangle$ representing the ground and the excited states of a single atom, and $\omega^{\xi}_n$  the energy of atom $\xi$ in  state $|n\rangle$.
The Hamiltonian of the fluctuating electromagnetic field reads
\bea
H_F(\tau)&=&\sum_{\nu}\int d^3\mathbf{k}\;\omega_{\mathbf{k}}a^{\dag}_{\mathbf{k},\nu}({t}(\tau))a_{\mathbf{k},\nu}({t}(\tau))\frac{d{t}}{d\tau}\;,
\label{Htot}
\eea
where $\nu$ is the polarization index, $\mathbf{k}$ stands for the wave vector, and $a^{\dag}_{\mathbf{k},\nu}$ and $a_{\mathbf{k},\nu}$ are the creation and annihilation operators, respectively. The interaction Hamiltonian  takes the form
\bea
H_I(\tau)&=&-\bm{\mu}^A(\tau)\cdot\bm{\mathrm{E}}(x_A(\tau))-\bm{\mu}^B(\tau)\cdot\bm{\mathrm{E}}(x_B(\tau))\;,
\label{Htot}
\eea
where $\bm{\mu}^\xi$ denotes the induced dipole moment of atom $\xi$, and $\bm{\mathrm{E}}(x)$ is the electric field operator,  given by
\bea
\bm{\mathrm{E}}(x)=\int d^3\mathbf{k}\sum\limits_{\nu=1}^2\frac{i\omega_{\mathbf{k}}\bm{\epsilon}(\mathbf{k},\nu)}{\sqrt{(2\pi)^32\omega_{\mathbf{k}}}}\left[a_{\mathbf{k},\nu}({t})e^{i\mathbf{k}\cdot \mathbf{x}}-a^{\dag}_{\mathbf{k},\nu}({t})e^{-i\mathbf{k}\cdot \mathbf{x}}\right]\label{field-operator}
\eea
where $\bm{\epsilon}(\mathbf{k},\nu)$ is the polarization vector satisfying the transversality condition~\cite{Greiner}:
\bea
\sum_{\nu=1}^2\epsilon_j(\mathbf{k},\nu)\epsilon_l(\mathbf{k},\nu)+\frac{k_jk_l}{\mathbf{k}^2}=\delta_{jl}\;.
\eea

Due to their interaction with the quantum field, an effective interaction potential may arise between the two atoms. To compute this potential, we employ the fourth-order Dalibard-Dupont-Roc-Cohen-Tannoudji (DDC) formalism~\cite{DDC82,DDC84}, following its recent extensions to higher-order perturbations~\cite{Zhou21,Cheng23}.

\subsection{The fourth-order DDC formalism.}

The Dalibard-Dupont-Roc-Cohen-Tannoudji (DDC) formalism, originally developed to second order for studying the evolution of a small quantum system interacting with a large reservoir~\cite{DDC82,DDC84}, has proven to be a powerful method for analyzing radiative properties of atoms arising from atom-field interactions. It has been extensively applied to investigate phenomena such as atomic transition rates~\cite{Meschede90,Audretsch94,Holzmann95,Tomazelli03,Yu05,Zhu06,Yu06,Zhou12}, radiative energy shifts of single atoms~\cite{Meschede90,Holzmann95,Audretsch95,Passante98,Rizzuto07,Rizzuto09,Zhu10}, and resonant interactions between entangled atomic pairs~\cite{Zhou16,Rizzuto16}. Since the interaction between two ground-state atoms involves fourth-order processes, the DDC formalism has recently been extended beyond second order~\cite{Zhou21,Cheng23}. Within this generalized framework, the resulting interatomic interaction potential can be separated into distinct contributions from field fluctuations and atomic radiation reactions.

For the two atoms in synchronized motion in vacuum, the separate contributions of field fluctuations $(\delta E)_{vf}$ and those of the atomic radiation reaction $(\delta E)_{rr}$ to the interatomic interaction potential are respectively given by~\cite{Cheng23}
\bea
(\delta E)_{vf}
&=&2i\sum_{j,l,k,h}\int_{\tau_0}^{\tau}d\tau_1\int_{\tau_0}^{\tau_1}d\tau_2\int_{\tau_0}^{\tau_2}d\tau_3\ C^F_{jl}(x_A(\tau),x_B(\tau_3))\chi^F_{kh}(x_A(\tau_1),x_B(\tau_2))\nn\\&&\times
\chi^A_{jk}(\tau,\tau_1)\chi^B_{hl}(\tau_2,\tau_3)+\text{$A\rightleftharpoons B$ {term}}
\label{vf-contriution}
\eea
and
\bea
&&(\delta E)_{rr}\nn\\
&=&2i\sum_{j,l,k,h}\int_{\tau_0}^{\tau}d\tau_1\int_{\tau_0}^{\tau_1}d\tau_2\int_{\tau_0}^{\tau_2}d\tau_3\ \chi^F_{jl}(x_A(\tau),x_B(\tau_3))\chi^F_{kh}(x_A(\tau_1),x_B(\tau_2))
C^A_{jk}(\tau,\tau_1)\chi^B_{hl}(\tau_2,\tau_3)\nonumber\\
&&+2i\sum_{j,l,k,h}\int_{\tau_0}^{\tau}d\tau_1\int_{\tau_0}^{\tau_1}d\tau_2\int_{\tau_0}^{\tau_2}d\tau_3\ \chi^F_{kl}(x_A(\tau_1),x_B(\tau_3))\chi^F_{jh}(x_A(\tau),x_B(\tau_2))
C^A_{jk}(\tau,\tau_1)\chi^B_{lh}(\tau_3,\tau_2)\nonumber\\
&&+2i\sum_{j,l,k,h}\int_{\tau_0}^{\tau}d\tau_1\int_{\tau_0}^{\tau_1}d\tau_2\int_{\tau_0}^{\tau_2}d\tau_3\ \chi^F_{hl}(x_B(\tau_2),x_A(\tau_3))\chi^F_{jk}(x_A(\tau),x_B(\tau_1))
C^A_{jl}(\tau,\tau_3)\chi^B_{kh}(\tau_1,\tau_2)\nonumber\\
&&+2i\sum_{j,l,k,h}\int_{\tau_0}^{\tau}d\tau_1\int_{\tau_0}^{\tau}d\tau_2\int_{\tau_0}^{\tau_2}d\tau_3\ \chi^F_{hl}(x_A(\tau_2),x_B(\tau_3))\chi^F_{jk}(x_A(\tau),x_B(\tau_1))
\chi^A_{jh}(\tau,\tau_2)C^B_{kl}(\tau_1,\tau_3)\nonumber\\
&&+2i\sum_{j,l,k,h}\int_{\tau_0}^{\tau}d\tau_1\int_{\tau_0}^{\tau_1}d\tau_2\int_{\tau_0}^{\tau}d\tau_3\ C^F_{lh}(x_B(\tau_2),x_A(\tau_3))\chi^F_{jk}(x_A(\tau),x_B(\tau_1))
\chi^A_{hj}(\tau_3,\tau)\chi^B_{lk}(\tau_2,\tau_1)\nonumber\\
&&+2i\sum_{j,l,k,h}\int_{\tau_0}^{\tau}d\tau_1\int_{\tau_0}^{\tau_1}d\tau_2\int_{\tau_0}^{\tau_1}d\tau_3\ \chi^F_{jl}(x_A(\tau),x_B(\tau_3))\chi^F_{kh}(x_A(\tau_1),x_B(\tau_2))
C^A_{jk}(\tau,\tau_1)\chi^B_{lh}(\tau_3,\tau_2)\nonumber\\
&&+\text{$A\rightleftharpoons B$ {terms}}\;,
\label{rr-contriution}
\eea
where $\tau_0$ denotes the onset time of the atom-field interaction. 
 The symmetric correlation function and linear susceptibility of the free field, $C^F_{jk}(x_{\xi}(\tau),x_{\xi'}(\tau'))$ and $\chi^F_{jk}(x_{\xi}(\tau),x_{\xi'}(\tau'))$, are defined as
\bea
C^F_{jk}(x_{\xi}(\tau),x_{\xi'}(\tau'))\equiv{1\/2}\langle 0|\left\{E^f_j(x_{\xi}(\tau)),E^f_k(x_{\xi'}(\tau'))\right\}|0\rangle\;,\quad\;\;
\label{CF}\\
\chi^F_{jk}(x_{\xi}(\tau),x_{\xi'}(\tau'))\equiv{1\/2}\langle 0|\left[E^f_j(x_{\xi}(\tau)),E^f_k(x_{\xi'}(\tau'))\right]|0\rangle \theta\left(\tau-\tau'\right)\;,
\label{chiF}
\eea
where $|0\rangle$ denotes the vacuum state, $\theta(x)$ is the Heaviside step function:
\bea
\theta(x)=\left\{
            \begin{array}{ll}
              1,\; x>0\;,\\
              0,\; x\leq0\;,
            \end{array}
          \right.
\eea
and the free field operator is given by
\beq
\bm{\mathrm{E}}^f(x)=\int d^3\mathbf{k}\sum\limits_{\nu=1}^{2}\frac{i\omega_{\mathbf{k}}\bm{\epsilon}(\mathbf{k},\nu)}{\sqrt{(2\pi)^3 2\omega_{\mathbf{k}}}}
    \left[a^f_{\mathbf{k},\nu}(t(\tau_0))e^{-i\omega_{\mathbf{k}}(t(\tau)-t(\tau_0))}e^{i\mathbf{k}\cdot\mathbf{x}}-H.c.\right]\;.
\eeq
The symmetric and antisymmetric statistical functions of atom $\xi$ are expressed as
\bea
C^{\xi}_{jk}(\tau,\tau')&\equiv&{1\/2}\langle g_{\xi}|\left\{\mathbf{\mu}_j^{\xi,f}(\tau),\mathbf{\mu}_k^{\xi,f}(\tau')\right\}|g_{\xi}\rangle
=(\mu^{\xi}_j)_{ge}(\mu^{\xi}_k)_{eg}\cos{[\omega_{\xi}(\tau-\tau')]}\;,\label{C-atom}\\
\chi^{\xi}_{jk}(\tau,\tau')&\equiv&{1\/2}\langle g_{\xi}|\left[\mathbf{\mu}_j^{\xi,f}(\tau),\mathbf{\mu}_k^{\xi,f}(\tau')\right]|g_{\xi}\rangle
=-i(\mu^{\xi}_j)_{ge}(\mu^{\xi}_k)_{eg}\sin{[\omega_{\xi}(\tau-\tau')]}\label{Chi-atom}
\eea
with $(\mu^{\xi}_j)_{ge}\equiv\langle g_{\xi}|\mu^{\xi}_{j}(\tau_0)|e_{\xi}\rangle$ 
and $\omega_{\xi}\equiv \omega^{\xi}_{e}-\omega^{\xi}_{g}$  the transition frequency of atom $\xi$.

The total interaction potential between the two atoms is then the sum of the vacuum fluctuation and radiation reaction contributions:
\beq
(\delta E)_{tot}=(\delta E)_{vf}+(\delta E)_{rr}\;.
\eeq

\subsection{The interatomic interaction potential.}\label{IIB}

To evaluate the contributions from vacuum fluctuations (vf-contribution) and atomic radiation reaction (rr-contribution), as given by Eqs.~(\ref{vf-contriution}) and (\ref{rr-contriution}), it is necessary to compute the symmetric correlation function and linear susceptibility of the electromagnetic field at two arbitrary points along the atomic trajectories specified in Eqs.~(\ref{trajectoryA}) and (\ref{trajectoryB}). To this end, we first derive the two-point correlation function of the electric field:
\bea
&&\langle0|E^f_j(x_A(\tau))E^f_k(x_B(\tau'))|0\rangle\nn\\
&=&\frac{a^4}{16\pi^2\left[\sinh^2\left(\frac{a}{2}\Delta\tau-i\epsilon\right)-\frac{1}{4}a^2L^2\right]^3}
\biggl\{\left[\delta_{jk}+aL\left(q_kn_j-q_jn_k\right)\right]\sinh^2\left(\frac{a}{2}\Delta\tau\right)\nn\\&&
+\frac{1}{4}a^2L^2\left(\delta_{jk}-2q_jq_k\right)\left[1+2\left(\delta_{jk}-n_jn_k\right)\sinh^2\left(\frac{a}{2}\Delta\tau\right)\right]\biggl\}
\label{instant-inertial-correlation-electric-field}
\eea
where $j, k$ denote spatial indices, $\Delta \tau=\tau-\tau'$, $\epsilon$ is a positive infinitesimal, and $\mathbf{n}=(1,0,0)$, and $\mathbf{q}=(0,0,1)$  indicate the directions of acceleration and interatomic separation, respectively.  By substituting this expression into Eqs.~(\ref{CF}) and (\ref{chiF}) and performing the Fourier transformation, we obtain
\bea
C^F_{jk}(x_A(\tau),x_B(\tau'))&=&-2\int_0^{\infty}d\omega'\left[f_{jk}(a,L)\omega'\cos(\omega' D_a)+g_{jk}(a,L,\omega')\sin(\omega' D_a)\right]\nn\\&&\times
\left(1+\frac{2}{e^{2\pi\omega'/a}-1}\right)\cos{(\omega'\Delta\tau)}\;,
\label{CF-acceleration}\\
\chi^F_{jk}(x_A(\tau),x_B(\tau'))&=&2i\int_0^{\infty}d\omega'\left[f_{jk}(a,L)\omega'\cos(\omega' D_a)+g_{jk}(a,L,\omega')\sin(\omega' D_a)\right]\nn\\&&\times
\sin{(\omega'\Delta\tau)}
\label{ChiF-acceleration}
\eea
with $D_a=\frac{2}{a}\sinh^{-1}{\left(\frac{1}{2}a L\right)}$, and $\mathcal{N}(a,L)=1+\frac{1}{4}a^2L^2$. The detailed expressions of $f_{jk}(a,L)$ and $g_{jk}(\omega,a,L)$ are given in Appendix~\ref{coefficient-functions}.

Now inserting Eqs.~(\ref{C-atom}), (\ref{Chi-atom}), (\ref{CF-acceleration}) and (\ref{ChiF-acceleration}) into Eqs.~(\ref{vf-contriution}) and (\ref{rr-contriution}), and then carrying out the triple-integrals over $\tau_1,\tau_2,\tau_3$ under the assumption that $\tau-\tau_0\rightarrow\infty$, and we arrive, after some algebraic simplifications, at  a compact expression for the interaction potential. It takes the following form:
\bea
(\delta E)_{q}&=&\sum_{jk}\alpha_j(A)\alpha_k(B)(\delta E)^{jk}_q\nn\\
&=&\alpha_x(A)\alpha_x(B)(\delta E)^{xx}_q+\alpha_y(A)\alpha_y(B)(\delta E)^{yy}_q+\alpha_z(A)\alpha_z(B)(\delta E)^{zz}_q\nn\\&&
+\alpha_x(A)\alpha_z(B)(\delta E)^{xz}_q+\alpha_z(A)\alpha_x(B)(\delta E)^{zx}_q\;,
\label{potential-general}
\eea
where $(\delta E)_q$, with $q=vf$, $rr$, or $tot$, correspond to the vf-contribution, the rr-contribution and the total interaction potential respectively, $\alpha_j(\xi)=\frac{2|\mu_j^{\xi}|^2}{3\omega_{\xi}}$, and
\bea
(\delta E)^{jk}_{vf}&=&18\pi\int^{\infty}_0d\omega_1\frac{\omega_A^2\omega_B^2\coth\left(\pi\omega_1/a\right)}{(\omega_1^2-\omega_A^2)(\omega_1^2-\omega_B^2)}
\big[f_{jk}(a,L)\omega_1\cos(\omega_1D_a)+g_{jk}(a,L,\omega_1)\sin(\omega_1D_a)\big]\nn\\&&\times
\left[f_{jk}(a,L)\omega_1\sin(\omega_1D_a)-g_{jk}(a,L,\omega_1)\cos(\omega_1D_a)\right]\;,
\label{acc-vf-contribution-1}
\eea
\bea
(\delta E)^{jk}_{rr}&=&36\pi\int^{\infty}_0d\omega_1\bigg[\frac{\omega_A\omega_B(\omega_1+\omega_A+\omega_B)}{(\omega_1+\omega_A)(\omega_1+\omega_B)(\omega_A+\omega_B)}
+\frac{2\omega^2_A\omega^2_B}{(\omega_1^2-\omega_A^2)(\omega_1^2-\omega_B^2)(e^{2\pi\omega_1/a}-1)}\bigg]\nn\\&&\times
[f_{jk}(a,L)\omega_1\cos(\omega_1D_a)+g_{jk}(a,L,\omega_1)\sin(\omega_1D_a)][f_{jk}(a,L)\omega_1\sin(\omega_1D_a)\nn\\&&
-g_{jk}(a,L,\omega_1)\cos(\omega_1D_a)]-(\delta E)^{jk}_{vf}\;,
\label{acc-rr-contribution-1}
\eea
and
\bea
(\delta E)^{jk}_{tot}&=&36\pi\int^{\infty}_0d\omega_1
\bigg[\frac{\omega_A\omega_B(\omega_1+\omega_A+\omega_B)}{(\omega_1+\omega_A)(\omega_1+\omega_B)(\omega_A+\omega_B)}
+\frac{2\omega^2_A\omega^2_B}{(\omega_1^2-\omega_A^2)(\omega_1^2-\omega_B^2)(e^{2\pi\omega_1/a}-1)}\bigg]\nn\\&&\times
[f_{jk}(a,L)\omega_1\cos(\omega_1D_a)+g_{jk}(a,L,\omega_1)\sin(\omega_1D_a)]
[f_{jk}(a,L)\omega_1\sin(\omega_1D_a)\nn\\&&
-g_{jk}(a,L,\omega_1)\cos(\omega_1D_a)]\;.\quad\quad
\label{acc-tot-contribution-1}
\eea

Equation (\ref{potential-general}) shows clearly  that both the diagonal components $(\delta E)_q^{jk}$ with $j=k$ and the off-diagonal components $(\delta E)_q^{jk}$ with $j\neq k$ 
contribute to the interaction potential. These off-diagonal terms, characterized by cross-polarizabilities such as $|\mu^A_x|^2_{ge}|\mu^B_z|^2_{ge}$ and $|\mu^A_z|^2_{ge}|\mu^B_x|^2_{ge}$, are a distinctive feature of the accelerated case. In contrast, such cross terms are absent for two inertial atoms in vacuum [see Eq. (50) in Ref.~\cite{Casimir-Polder48}] as well as for two static atoms in a thermal bath [see Eqs. (50) and (53)-(55) in Ref.~\cite{Cheng23}]. This implies that the emergence of off-diagonal contributions is a direct consequence of the acceleration. In particular, when two cross-polarizable atoms are considered, i.e., atoms respectively polarizable along the direction of separation and a perpendicular direction (e.g., the direction of acceleration), a nonvanishing interatomic interaction arises only under acceleration. Such an interaction is absent both for inertial atoms in vacuum and for static atoms in a thermal bath, consistent with the findings of Ref.~\cite{Zhou25}. Therefore, the emergence of an interaction between two cross-polarizable atoms is a distinctive signal of the FDU effect, much like the spontaneous excitation of a single accelerated atom. This interaction force, which acts parallel to the interatomic separation and may be either attractive or repulsive~\cite{Zhou25}, leads to a measurable contraction or expansion of the separation between the atoms. Such behavior in principle offers a promising route toward experimental verification of the FDU effect: both the reduction and the enlargement of the interatomic distance constitute clear signatures of the underlying acceleration-induced quantum field effects.

To proceed with a more detailed analysis, we now focus on the specific behavior of the interaction potential. For simplicity, we assume the two atoms are identical, such that $\omega_A=\omega_B\equiv \omega$ or equivalently $\lambda_{\xi}\equiv\lambda=2\pi\omega^{-1}$.

\section{Acceleration effects on the interatomic interaction potential}

Before delving into  the effects of acceleration on the interatomic interaction potential, 
it is helpful to recall the well-established behavior of the potential in the inertial case.   In that scenario, the potential exhibits two distinct regimes: in the van der Waals (vdW) region, where the interatomic separation $L$ is much smaller than $\lambda$, it scales as $\sim L^{-6}$~\cite{London30};  in the Casimir-Polder (CP) region where $L\gg\lambda$, the effect of retardation modifies the scaling to $\sim L^{-7}$~\cite{Casimir-Polder48}.

In the accelerated case, an additional length scale emerges: $L_a=a^{-1}$, which characterizes the breakdown of the local inertial frame approximation. As a result, the conventional vdW and CP regions are further subdivided. Specifically, the vdW region  $L\ll\lambda$ is partitioned into three subregions: $L\ll\lambda\ll L_a$, $L\ll L_a\ll\lambda$ and $ L_a\ll L\ll\lambda$, and we label them by I, II and III, respectively.
The CP region  $L\gg\lambda$ is also divided into: $\lambda\ll L\ll L_a$, $\lambda\ll L_a\ll L$ and $ L_a\ll\lambda\ll L$, and we label them by IV, V and VI, respectively.

\subsection{Van der Waals interaction potential.}

\subsubsection{VdW interaction potential in region I where $L\ll\lambda\ll L_a$.} 

Table~\ref{tab1} summarizes the vdW interaction potential and its vacuum fluctuation (vf) and radiation reaction (rr) contributions in Region I. Hereafter, $(\delta E)^{jk}_{tot}$ denotes the sum of $(\delta E)^{jk}_{vf}$ and $(\delta E)^{jk}_{rr}$. 
\begin{center}
\begin{table}[H]
	\renewcommand{\arraystretch}{1.1}
    \caption{VdW interaction potential in region I where $L\ll\lambda\ll L_a$.}
\begin{threeparttable}
\begin{tabular}{|c|c|c|c|}
   \hline        $jk$    & $(\delta E)^{jk}_{vf}$ & $(\delta E)^{jk}_{rr}$ &  $(\delta E)^{jk}_{tot}$ \\
  \hline $xx$   & \multirow{2}{*}{$-\frac{3\omega^4\ln(2\omega L)}{16\pi^3L^3}+\frac{(23-24\gamma)\omega^4}{128\pi^3L^3}-\frac{3a^2\omega^2}{32\pi^3L^3}$}
                                                 & $-\frac{9\omega}{128\pi^2L^6}-\frac{9\omega^3}{128\pi^2L^4}-\frac{3\omega^4\ln(2\omega L)}{16\pi^3L^3}+\frac{9a^2\omega }{512\pi^2 L^4}$                                                 &  \multirow{2}{*}{$-\frac{9\omega}{128\pi^2L^6}
                                                 $}\\
\cline{1-1} \cline{3-3} $yy$                     &
                                                 & $-\frac{9\omega}{128\pi^2L^6}-\frac{9\omega^3}{128\pi^2L^4}-\frac{3\omega^4\ln(2\omega L)}{16\pi^3L^3}+\frac{27a^2\omega }{512\pi^2 L^4}$                                                 &  \\
\hline \multirow{2}{*}{$zz$}        & \multirow{2}{*}{$\frac{3\omega^4\ln(2\omega L)}{8\pi^3L^3}+\frac{(-11+6\gamma)\omega^4}{16\pi^3L^3}+\frac{3a^2\omega^2}{16\pi^3L^3}$}
                                    & $-\frac{9\omega}{32\pi^2L^6}+\frac{9\omega^3}{32\pi^2L^4}+\frac{3\omega^4\ln(2\omega L)}{8\pi^3L^3}$
                                    & \multirow{2}{*}{$-\frac{9\omega}{32\pi^2L^6}$}\\
                                    &
                                    & $+\frac{(-11+6\gamma)\omega^4}{16\pi^3L^3}+\frac{3a^2\omega^2}{16\pi^3 L^3}$
                                    & \\
 \hline $xz$ or $zx$         & $\frac{3a^2\omega^4\ln{(2\omega L)}}{32\pi^3L}$
                             & $-\frac{9a^2\omega}{512\pi^2L^4}$
                             & $-\frac{9a^2\omega}{512\pi^2L^4}$\\
\hline
\end{tabular}
\end{threeparttable}
\label{tab1}
\end{table}
\end{center}

We can see from Tab.~\ref{tab1} that the diagonal components $(\delta E)_{vf,rr}^{jk}$ with $j=k=x, z, y$, corresponding to the acceleration direction, the interatomic separation direction, and the transverse direction, are dominated by acceleration-independent terms. Acceleration-dependent terms appear only at higher order and thus produce negligible corrections. In contrast, the leading-order off-diagonal components $(\delta E)_{vf,rr}^{xz}$ and $(\delta E)_{vf,rr}^{zx}$ are crucially acceleration-dependent, which is consistent with our earlier analysis: these off-diagonal terms arise exclusively due to acceleration. Furthermore, comparing the vf- and rr-contributions reveals that $|(\delta E)_{vf}^{jk}|\ll|(\delta E)_{rr}^{jk}|$,  indicating that the total interaction is overwhelmingly dominated by the rr-contribution. Notably, the leading-order terms of $(\delta E)_{rr}^{jk}$ are all negative, with diagonal components scaling as $\sim L^{-6}$, and the off-diagonal components $(\delta E)_{rr}^{xz}$ and $(\delta E)_{rr}^{zx} \sim a^2L^{-4}$.

Since components of the interaction potential $(\delta E)_q^{jk}$ (with $q=vf, rr$ or $tot$) generally differ, the atomic polarizability plays a very important role in shaping the interaction.  In particular, for two isotropically polarizable atoms, the dominant rr-contribution yields the following total interaction potential\footnote{Hereafter, the interaction potential between two isotropically polarizable atoms is expressed in units of $\alpha(A)\alpha(B)$ where $\alpha(\xi)=\sum\limits_j\alpha_j(\xi)$ denotes the polarizability of atom $\xi$.}:
\beq
(\delta E)^{iso}_{tot}\approx-\frac{3\omega}{64\pi^2L^6}\;,
\eeq
which coincides with  the leading-order vdW interaction between two isotropically polarizable inertial atoms in vacuum~\cite{London30}. By contrast,  the interaction potential between two cross-polarizable atoms in this region scales as $\sim a^2L^{-4}$ (see the last row in Tab.~\ref{tab1}), vanishing in the inertial limit
$a\rightarrow0$.

\subsubsection{VdW interaction potential in region II where $L\ll L_a\ll\lambda$.}

The approximate analytical results for the interaction potential in the second vdW subregion $L\ll L_a\ll\lambda$ are displayed in Tab.~\ref{tab2}.
\begin{center}
\begin{table}[H]
	\renewcommand{\arraystretch}{1.2}
    \caption{VdW interaction potential in region II where $L\ll L_a\ll\lambda$.}
\begin{threeparttable}
\begin{tabular}{|c|c|c|c|}
   \hline      $jk$   & $(\delta E)^{jk}_{vf}$ & $(\delta E)^{jk}_{rr}$ &  $(\delta E)^{jk}_{tot}$ \\
  \hline $xx$         & \multirow{2}{*}{$\frac{3a^2\omega^2}{32\pi^3L^3}$}
                                                 & $-\frac{9\omega}{128\pi^2L^6}+\frac{9a^2\omega}{512\pi^2L^4}$                                                 &  \multirow{2}{*}{$-\frac{9\omega}{128\pi^2L^6}
                                                $}\\
  \cline{1-1} \cline{3-3} $yy$                     &
                                                 & $-\frac{9\omega}{128\pi^2L^6}+\frac{27a^2\omega}{512\pi^2L^4}
                                                 $                                                 &  \\
  \hline $zz$         & $-\frac{3a^2\omega^2}{16\pi^3L^3}$
                      & $-\frac{9\omega}{32\pi^2L^6}+\frac{9\omega^3}{32\pi^2L^4}-\frac{3a^2\omega^2}{16\pi^3L^3}$                                                 &  $-\frac{9\omega}{32\pi^2L^6}
                                                 $\\
 \hline $xz$ or $zx$  & $-\frac{3a^4\omega^2}{64\pi^3L}$
                      & $-\frac{9a^2\omega}{512\pi^2L^4}
                                                 $
                      & $-\frac{9a^2\omega}{512\pi^2L^4}$ \\
\hline
\end{tabular}
\end{threeparttable}
\label{tab2}
\end{table}
\end{center}

The results in Table~\ref{tab2} reveal several key differences compared to Region I. First,  the diagonal components $(\delta E)_{vf}^{jk}$ with $j=k$,  unlike in Region I, where they  are subleading, scale as $\sim a^2L^{-3}$, indicating a pronounced dependence on acceleration. Interestingly, their magnitude matches that in Region I, but the sign is reversed.  Second, the off-diagonal components  $(\delta E)_{vf}^{jk}$ with $j\neq k$ transition from a $\sim a^2L^{-1}\ln{(2\omega L)}$ behavior in region I to the $\sim a^4L^{-1}$ behavior in region II,  reflecting a more rapid decay and a higher-order dependence on acceleration. Third,  diagonal components $(\delta E)_{rr}^{jk}$ with $j=k$, while still dominated by acceleration-independent terms, now include lower-order acceleration-dependent corrections compared to Region I.

These observations suggest that acceleration plays a more substantial role in the interatomic interaction in Region II than in Region I. Nevertheless, the leading-order total interaction potential remains unchanged between the two regions (compare the last columns of Tables~\ref{tab1} and \ref{tab2}). This is because the dominant contributions to the diagonal and off-diagonal rr-components are the same in both regions and still outweigh those of the vf-contribution.

\subsubsection{VdW interaction potential in region III where $L_a\ll L\ll\lambda$.}
As shown in Tab.~\ref{tab3}, the interaction potential in the third vdW subregion $L_a\ll L\ll\lambda$ is significantly modified by the accelerated motion, resulting in completely new behaviors markedly distinctive from those in regions I and II.
\begin{center}
\begin{table}[H]
	\renewcommand{\arraystretch}{1.2}
    \caption{VdW interaction potential in region III where $L_a\ll L\ll \lambda$.}
\begin{threeparttable}
\begin{tabular}{|c|c|c|c|}
   \hline   $jk$  & $(\delta E)^{jk}_{vf}$ & $(\delta E)^{jk}_{rr}$ &  $(\delta E)^{jk}_{tot}$ \\
  \hline $xx$                    & $-\frac{18\omega^2\ln{(a L)}}{\pi^3a^3L^8}$
                                 & $-\frac{9\omega}{2\pi^2a^2L^8}$
                                 & $-\frac{9\omega}{2\pi^2a^2L^8}$\\
  \hline $yy$                    & $-\frac{9\omega^4\ln{(a L)}\ln{(\frac{a^2L}{\omega})}}{4\pi^3aL^4}+\frac{9\omega^2}{4\pi^3a^3L^8}$
                                 & $-\frac{9\omega^3}{32\pi^2L^4}+\frac{9\omega^2}{4\pi^3a^3L^8}$
                                 & $-\frac{9\omega^3}{32\pi^2L^4}+\frac{9\omega^2}{2\pi^3a^3L^8}$ \\
  \hline $zz$                    & $-\frac{9\omega^4\ln{(a L)}\ln{(\frac{a^2L }{\omega})}}{4\pi^3aL^4}+\frac{45\omega^2}{4\pi^3a^3L^8}$
                                 & $-\frac{9\omega^3}{32\pi^2L^4}+\frac{45\omega^2}{4\pi^3a^3L^8}$
                                 & $-\frac{9\omega^3}{32\pi^2L^4}+\frac{45\omega^2}{2\pi^3a^3L^8}$ \\
  \hline $xz$ or $zx$            & $-\frac{9\omega^4\ln{(a L)}\ln{(\frac{a^2L}{\omega})}}{\pi^3a^3L^6}-\frac{9\omega^2}{\pi^3 a^3 L^8}$
                                 & $-\frac{9\omega^3}{8\pi^2a^2L^6}-\frac{9\omega^2}{\pi^3 a^3 L^8}$
                                 & $-\frac{9\omega^3}{8\pi^2a^2L^6}-\frac{18\omega^2}{\pi^3 a^3 L^8}$ \\
  \hline
\end{tabular}
\end{threeparttable}
\label{tab3}
\end{table}
\end{center}

Several notable features emerge in Region III.  The first term in $(\delta E)^{yy}_{vf}$ and $(\delta E)^{zz}_{vf}$ is not assuredly larger than the second one, and the coefficient of the second term in  $(\delta E)^{yy}_{vf}$ and $(\delta E)^{zz}_{vf}$ differs, indicating nuanced acceleration-induced asymmetries. A similar distinction exists between  $(\delta E)^{yy}_{rr}$ and $(\delta E)^{zz}_{rr}$. Second, the leading-order term in  $(\delta E)^{xx}_{vf}$ differs significantly from those in $(\delta E)^{yy}_{vf}$ and $(\delta E)^{zz}_{vf}$, {a departure from the behavior in Regions I and II where  $(\delta E)^{xx}_{vf} = (\delta E)^{yy}_{vf} \neq (\delta E)^{zz}_{vf}$}.  Similar changes also appear in the rr-contribution.  Third,  while the leading  $ L^{-4}$ term in  $(\delta E)^{yy}_{rr}$ and $(\delta E)^{zz}_{rr}$ may appear acceleration-independent, it in fact reflects acceleration-induced modifications. This scaling is distinct from the $ L^{-6}$   behavior in both the inertial case and Regions I and II.

These observations indicate that both vf- and rr-contributions are heavily modified in this regime. As seen in the final column of Table~\ref{tab3}, the total interaction potential features terms scaling as $\sim L^{-4}$, $a^{-2}L^{-6}$, $a^{-2}L^{-8}$ and $a^{-3}L^{-8}$, depending on the atomic polarizabilities. For example, the first term $-\frac{9\omega^3}{32\pi^2L^4}$ in $(\delta E)^{yy}_{tot}$ and $(\delta E)^{zz}_{tot}$ dominates when  $\sqrt[4]{\lambda L_a^3}\ll L\ll\lambda$, while the second term $\sim a^{-3}L^{-8}$ dominates when $ L_a\ll L\ll\sqrt[4]{\lambda L_a^3}$. Similarly, in $(\delta E)^{xz}_{tot}$ and $(\delta E)^{zx}_{tot}$, the first term $\sim a^{-2}L^{-6}$ dominates for $\sqrt{\lambda L_a}\ll L\ll\lambda$, whereas the second $\sim a^{-3}L^{-8}$ prevails for $L_a\ll L\ll\sqrt{\lambda L_a}$.

The total interaction potential depends on the full contraction $\sum\limits_{j,k}\alpha_j(A)\alpha_k(B)(\delta E)^{jk}_{tot}$. For illustration, consider the case of two isotropically polarizable atoms. The interaction potential, dominated by the rr-contribution, is
\beq
(\delta E)^{iso}_{tot}\approx(\delta E)^{iso}_{rr}\approx-\frac{\omega^3}{16\pi^2L^4}-\frac{\omega}{2\pi^2a^2L^8}\;.
\label{iso-III}
\eeq
Depending on the range of $L$, Region III can be further subdivided into: subregion III-$a$ where $L_a\ll L\ll\sqrt{\lambda L_a}$  and subregion III-$b$ where $\sqrt{\lambda L_a}\ll L\ll\lambda$. In subregion III-$a$, the second term dominates, giving
 \beq
(\delta E)^{iso}_{tot}\approx-\frac{\omega}{2\pi^2a^2L^8}\;,
\eeq
primarily due to $(\delta E)^{xx}_{tot}$. In subregion III-$b$, the first term dominates, leading to
\beq
(\delta E)^{iso}_{tot}\approx-\frac{\omega^3}{16\pi^2L^4}\;,\label{iso-III-second}
\eeq
equally coming from $(\delta E)^{yy}_{tot}$ and $(\delta E)^{zz}_{tot}$.

Having now examined the interaction potential across all three vdW subregions, $L\ll\lambda\ll L_a$, $L\ll L_a\ll\lambda$ and $ L_a\ll L\ll\lambda$,  we summarize the acceleration-induced effects.  First, these regions respectively correspond to small, moderate, and large accelerations for fixed  $L\ll\lambda$. Second, acceleration effects are negligible in Region I, more significant in Region II (though the leading-order interaction remains unaltered  as compared with that of two inertial atoms), and most pronounced in Region III, where novel behaviors emerge. Third, depending on the atomic polarizability, separation, and acceleration, acceleration can either enhance or suppress the vdW interaction. For instance, in Region II, when both atoms are polarizable only along the interatomic axis, the interaction is enhanced due to the dominant vf-contribution  $-\frac{3a^2\omega^2}{16\pi^3L^3}$ and the strengthened rr-term $(\delta E)_{rr}^{zz}$ (see the fourth row of Tab.~\ref{tab2}). This results in an overall enhancement of the interaction potential by $-\frac{3a^2\omega^2}{8\pi^3L^3}$, even though the scaling $\sim L^{-6}$  of the leading order remains unchanged.

\subsection{Casimir-Polder interaction potential.}
We now turn our attention to the interaction potential in the three CP subregions, i.e., regions IV, V, and VI, where $\lambda\ll L\ll L_a$, $\lambda\ll L_a\ll L$, and $L_a\ll\lambda\ll L$, respectively.

\subsubsection{CP interaction potential in region IV where $\lambda\ll L\ll L_a$.}
The results for the interaction potential in the first CP subregion $\lambda\ll L\ll L_a$ are summarized in Tab.~\ref{tab4}, where $g_s(\omega L)$  represents a series of dimensionless functions, and for the detailed expressions of these functions, see Appendix~\ref{gh}. Hereafter, the upper and lower signs before the oscillatory terms apply to $(\delta E)^{jk}_{vf}$ and $(\delta E)^{jk}_{rr}$, respectively.
\begin{center}
\begin{table}[H]
	\renewcommand{\arraystretch}{1.2}
    \caption{CP interaction potential in region IV where $\lambda\ll L\ll L_a$.}
\begin{threeparttable}
\begin{tabular}{|c|c|c|c|}
   \hline         $jk$ & $(\delta E)^{jk}_{vf},\;(\delta E)^{jk}_{rr}$ &  $(\delta E)^{jk}_{tot}$ \\
  \hline
\multirow{2}{*}{$xx$}  & $\pm\frac{9\omega^6\left[g_1(\omega L)-\frac{3}{4}a^2L^2g_2(\omega L)\right]\sin(2\omega L)}{128\pi^2L}\pm\frac{9\omega^5\left[g_1(\omega L)-\frac{1}{6}a^2L^4\omega^2g_3(\omega L)\right]\cos(2\omega L)}{256\pi^2L^2}$
                       & \multirow{2}{*}{$-\frac{117}{128\pi^3L^7}-\frac{9a^2}{1024\pi^3L^5}$} \\
                       & $-\frac{117}{256\pi^3L^7}-\frac{9a^2}{2048\pi^3L^5}\quad\quad\quad\quad\quad\quad\quad\quad\quad\quad\quad\quad\quad\quad\quad\quad\quad\quad\quad\quad\;\;$
                       & \\
  \hline
\multirow{2}{*}{$yy$}  & $\pm\frac{9\omega^6\left[g_1(\omega L)-\frac{1}{4}a^2L^2g_4(\omega L)\right]\sin(2\omega L)}{128\pi^2L}\pm\frac{9\omega^5\left[g_1(\omega L)-\frac{1}{6}a^2L^4\omega^2g_5(\omega L)\right]\cos(2\omega L)}{256\pi^2L^2}$
                       & \multirow{2}{*}{$-\frac{117}{128\pi^3L^7}+\frac{63a^2}{1024\pi^3L^5}$} \\
                       & $-\frac{117}{256\pi^3L^7}+\frac{63a^2}{2048\pi^3L^5}\quad\quad\quad\quad\quad\quad\quad\quad\quad\quad\quad\quad\quad\quad\quad\quad\quad\quad\quad\quad\;\;$
                       & \\
  \hline
\multirow{2}{*}{$zz$}  & $\mp\frac{9\omega^4\left[g_6(\omega L)-\frac{2}{3}a^2L^2\right]\sin(2\omega L)}{32\pi^2L^3}\mp\frac{27\omega^3\left[g_7(\omega L)-\frac{2}{9}a^2L^4\omega^2g_8(\omega L)\right]\cos(2\omega L)}{64\pi^2L^4}\quad\quad$
                       & \multirow{2}{*}{$-\frac{45}{32\pi^3L^7}+\frac{9a^2}{64\pi^3L^5}$} \\
                       & $-\frac{45}{64\pi^3L^7}+\frac{9a^2}{128\pi^3L^5}\quad\quad\quad\quad\quad\quad\quad\quad\quad\quad\quad\quad\quad\quad\quad\quad\quad\quad\quad\quad\quad\;\;$
                       & \\
  \hline
\multirow{1}{*}{$xz$ or $zx$}
                       & $\pm\frac{9a^2\omega^6Lg_1(\omega L)\sin(2\omega L)}{512\pi^2}\mp\frac{63a^2\omega^5g_9{(\omega L)}\cos(2\omega L)}{1024\pi^2}-\frac{27a^2}{1024\pi^3L^5}$
                       & \multirow{1}{*}{$-\frac{27a^2}{512\pi^3L^5}$} \\
   \hline
\end{tabular}
\end{threeparttable}
\label{tab4}
\end{table}
\end{center}

From the second column of Tab.~\ref{tab4}, we  observe that both the vf- and rr-contributions to the interaction potential, $(\delta E)^{jk}_{vf}$ and $(\delta E)^{jk}_{rr}$, contain oscillatory and nonoscillatory terms. The amplitudes of the oscillatory terms are significantly larger than those of the non-oscillatory terms, causing the interaction potential to oscillate with the interatomic separation $L$. These oscillations can take positive or negative values, or even cancel each other out. In contrast, the total interaction potential, $(\delta E)^{jk}_{tot}$, which includes both contributions, scales monotonically with $L$, because the oscillatory terms from the vf and rr contributions effectively cancel each other out. The diagonal components $(\delta E)^{jk}_{tot}$ with $j=k$ are dominated by acceleration-independent terms that scale as $L^{-7}$, with acceleration-dependent terms only contributing as small modifications. These acceleration-dependent modifications have the same scaling behavior as the leading-order terms of the off-diagonal components, $(\delta E)^{xz}_{tot}$ and $(\delta E)^{zx}_{tot}$. Comparing the total interaction potential in the last column of Table~\ref{tab4} with the corresponding terms in Table~\ref{tab1}, we find that the separation-dependence of the leading-order CP interaction potential in this region ($\lambda\ll L\ll  L_a$) scales as $ L^{-7}$ for the diagonal components and $L^{-5}$ for the off-diagonal components. This behavior represents an order of magnitude higher than the vdW interaction potential in Region I, where $L\ll\lambda\ll L_a$, which scales as $ L^{-6}$ for the diagonal components and $L^{-4}$ for the off-diagonal components, respectively. This difference is a result of retardation effects.

When both atoms are isotropically polarizable, the vacuum fluctuation and radiation reaction contributions combine to yield the total interaction potential:
\beq
(\delta E)^{iso}_{tot}\approx-\frac{23}{64\pi^3L^7}+\frac{5a^2}{512\pi^3L^5}\;.
\label{iso-IV}
\eeq
This expression indicates that the acceleration effects in this region are negligible, as the interaction potential scales in the leading order monotonically with the interatomic separation as $\sim L^{-7}$, the same as the interaction between two inertial atoms in vacuum~\cite{Casimir-Polder48}.


\subsubsection{CP interaction potential in region V where $\lambda\ll L_a\ll L$.}
Similar to the first CP subregion $\lambda\ll L\ll L_a$, both the  vf-contribution $(\delta E)^{jk}_{vf}$ and the rr-contribution $(\delta E)^{jk}_{rr}$ in the second CP subregion $\lambda\ll L_a\ll L$ contain both oscillatory and the nonoscillatory terms.  The absolute values of the amplitudes of the oscillatory terms  are much greater than those of nonoscillatory terms (see Tab.~\ref{tab5}~\footnote{For explicit expressions of the dimensionless functions $h_s(\frac{a}{\omega},aL)$ with $s$ ranging from $1$ to $16$, refer to Appendix~\ref{gh}.}). Consequently, the total interaction potential equally coming from the vf- and rr-contribution scales monotonically with the interatomic separation, as the oscillatory terms  in both contributions cancel each other out perfectly.  Despite these similarities with Region IV,  the interaction potential in this region differs  significantly in that the diagonal terms $(\delta E)_{jk}^{tot}$ with $j=k$ are no longer dominated by acceleration-independent terms.   Instead, they are crucially dependent on acceleration, scaling as $\sim a^{-1}L^{-8}$. In contrast, the off-diagonal components $(\delta E)^{jk}_{tot}$ with $j\neq k$  display an $a^{-3}L^{-10}$-dependence,  which is clearly distinctive from the $a^2L^{-5}$-dependence  in region IV (see Tab.~\ref{tab4}). Since $a^{-3}L^{-10}\ll a^{-1}L^{-8}$ in this region, the total interaction potential of two isotropically polarizable atoms is approximately given by
\beq
(\delta E)^{iso}_{tot}\approx-\frac{2}{\pi^3aL^8}\;.\label{V-iso}
\eeq
This expression stands in stark contrast to that in the first CP subregion $\lambda\ll L\ll L_a$ (see Eq.~(\ref{iso-IV})).

\begin{center}
\begin{table}[H]
	\renewcommand{\arraystretch}{1.2}
    \caption{CP interaction potential in region V where $\lambda\ll L_a\ll L$.}
\begin{threeparttable}
\begin{tabular}{|c|c|c|}
   \hline         $jk$   & $(\delta E)^{jk}_{vf},\;(\delta E)^{jk}_{rr}$ &   $(\delta E)^{jk}_{tot}$ \\
  \hline
  \multirow{1}{*}{$xx$}  & $\pm\frac{36\omega^5\ln{(aL)}h_1(\frac{a}{\omega},aL)}{\pi^2a^6L^8}\cos(\frac{4\omega}{a}\ln(aL))\pm\frac{9\omega^6\ln{(aL)}h_2(\frac{a}{\omega},aL)}{\pi^2a^7L^8}\sin(\frac{4\omega}{a}\ln(aL))-\frac{9}{2\pi^3aL^8}$
                         &  \multirow{1}{*}{$-\frac{9}{\pi^3aL^8}$} \\
\hline
  \multirow{2}{*}{$yy$}                      & $\pm\frac{9\omega^5\ln{(aL)}\left[h_3(\frac{a}{\omega},aL)-\frac{4}{a^2L^2}h_4(\frac{a}{\omega},aL)+\frac{18}{a^4L^4}h_5(\frac{a}{\omega},aL)\right]}{8\pi^2a^2L^4}\cos(\frac{4\omega}{a}\ln(aL))\quad\quad\quad\quad$
  &  \multirow{4}{*}{$-\frac{9}{2\pi^3aL^8}$}\\
  & $\pm\frac{9\omega^6\ln{(aL)}\left[h_6(\frac{a}{\omega},aL)-\frac{4}{a^2L^2}h_7(\frac{a}{\omega},aL)+\frac{16}{a^4L^4}h_8(\frac{a}{\omega},aL)\right]}{16\pi^2a^3L^4}\sin(\frac{4\omega}{a}\ln(aL))-\frac{9}{4\pi^3aL^8}$
  & \\
\cline{1-2}
\multirow{2}{*}{$zz$}
  & $\pm\frac{9\omega^5\ln{(aL)}\left[h_3(\frac{a}{\omega},aL)-\frac{12}{a^2L^2}h_{9}(\frac{a}{\omega},aL)+\frac{114}{a^4L^4}h_{10}(\frac{a}{\omega},aL)\right]}{8\pi^2a^2L^4}\cos(\frac{4\omega}{a}\ln(aL))\quad\quad\quad\quad$
  & \\
  & $\pm\frac{9\omega^6\ln{(aL)}\left[h_6(\frac{a}{\omega},aL)-\frac{12}{a^2L^2}h_{11}(\frac{a}{\omega},aL)+\frac{96}{a^4L^4}h_{12}(\frac{a}{\omega},aL)\right]}{16\pi^2a^3L^4}\sin(\frac{4\omega}{a}\ln(aL))-\frac{9}{4\pi^3aL^8}$
  & \\
   \hline \multirow{2}{*}{$xz$ or $zx$}
  & $\pm\frac{9\omega^5\ln{(aL)}\left[h_{3}(\frac{a}{\omega},aL)-\frac{16}{a^2L^2}h_{13}(\frac{a}{\omega},aL)+\frac{154}{a^4L^4}h_{14}(\frac{a}{\omega},aL)\right]}{2\pi^2a^4L^6}\cos(\frac{4\omega}{a}\ln(aL))\quad\quad\quad\quad$
  &  \multirow{2}{*}{$-\frac{54}{\pi^3a^3L^{10}}$} \\
  & $\pm\frac{9\omega^6\ln{(aL)}\left[h_6(\frac{a}{\omega},aL)-\frac{12}{a^2L^2}h_{15}(\frac{a}{\omega},aL)+\frac{96}{a^4L^4}h_{16}(\frac{a}{\omega},aL)\right]}{4\pi^2a^5L^6}\sin(\frac{4\omega}{a}\ln(aL))-\frac{27}{\pi^3a^3L^{10}}$
  & \\
   \hline
\end{tabular}
\end{threeparttable}
\label{tab5}
\end{table}
\end{center}

\subsubsection{CP interaction potential in region VI where $L_a\ll\lambda\ll L$.}
The interaction potential of the two accelerated atoms in the third CP subregion $L_a\ll\lambda\ll L$ is shown in the following Tab.~\ref{tab6}.
\begin{center}
\begin{table}[H]
	\renewcommand{\arraystretch}{1.2}
    \caption{CP interaction potential in region VI where $L_a\ll\lambda\ll L$.}
\begin{threeparttable}
\begin{tabular}{|c|c|c|c|}
   \hline         $jk$   & $(\delta E)^{jk}_{vf}$ & $(\delta E)^{jk}_{rr}$ &  $(\delta E)^{jk}_{tot}$ \\
  \hline $xx$                    & $-\frac{18\omega^2\ln{(a L)}}{\pi^3a^3L^8}$
                                 & $-\frac{9\omega}{2\pi^2a^2L^8}$
                                 &  $-\frac{9\omega}{2\pi^2a^2L^8}$ \\
  \hline $yy$                    & \multirow{2}{*}{$-\frac{9\omega^4\ln{(a L)}\ln{(\frac{a^2L}{\omega})}}{4\pi^3aL^4}$}
                                 & \multirow{2}{*}{$-\frac{9\omega^3}{32\pi^2L^4}$ }
                                 & \multirow{2}{*}{$-\frac{9\omega^3}{32\pi^2L^4}$}\\
  \cline{1-1}  $zz$              & 
                                 & 
                                 & 
                                   \\
  \hline $xz$ or $zx$            & $-\frac{9\omega^4\ln{(a L)}\ln{(\frac{a^2L}{\omega})}}{\pi^3a^3L^6}$
                                 & $-\frac{9\omega^3}{8\pi^2a^2L^6}$
                                 & $-\frac{9\omega^3}{8\pi^2a^2L^6}
                                        $ \\
  \hline
\end{tabular}
\end{threeparttable}
\label{tab6}
\end{table}
\end{center}

In sharp contrast to the results in the previous  two CP subregions $\lambda\ll L\ll L_a$ and $\lambda\ll L_a\ll L$, both the vf- and rr-contributions $(\delta E)^{jk}_{vf}$ and $(\delta E)^{jk}_{rr}$ in this third CP region no longer exhibit oscillatory behavior with respect to the interatomic separation $L$. Instead, they scale monotonically with $L$ and no longer contribute equally to the total interaction potential $(\delta E)^{jk}_{tot}$ because now $|(\delta E)^{jk}_{vf}|\ll|(\delta E)^{jk}_{rr}|$. Notably, the dominant rr-contribution leads to an  $a^{-2}L^{-8}$-dependence for $(\delta E)^{xx}_{tot}$, and an $a^{-2}L^{-6}$-dependence for $(\delta E)^{xz}_{tot}$ and $(\delta E)^{z x}_{tot}$, which are clearly acceleration-dependent. In contrast, the dominant rr-contribution leads to  an $a^0L^{-4}$-dependence for $(\delta E)^{yy}_{tot}$ and $(\delta E)^{zz}_{tot}$, which is not explicitly acceleration-dependent.  Recalling that the CP interaction potential of two inertial atoms scales as  $\sim L^{-7}$ in vacuum~\cite{Casimir-Polder48}, we see that  the $(\delta E)^{yy}_{tot}$ and $(\delta E)^{zz}_{tot}$ in this region are actually significantly altered by the accelerated motion.  These alterations are distinct from those seen in the second CP subregion $\lambda\ll L_a\ll L$ (see the last column of this table and their counterparts in Tab.~\ref{tab5}). As a result, the interaction potential for two isotropically polarizable atoms in this region  manifests a new behavior:
\beq
(\delta E)^{iso}_{tot}\approx-\frac{\omega^3}{16\pi^2L^4}\;.
\eeq
Interestingly, this expression of the interaction potential in region VI, where $L_a\ll\lambda\ll L$, is identical to that in the furthest vdW subregion where $\sqrt{\lambda L_a}\ll L\ll\lambda$ (see Eq.~(\ref{iso-III-second})). Both regions, $L_a\ll\lambda\ll L$ and $\sqrt{\lambda L_a}\ll L\ll\lambda$, share the condition $L\gg\frac{\lambda}{L}L_a$. This suggests that, in the acceleration case, both the vdW and CP interaction potentials for two isotropically polarizable atoms, which display obviously distinctive behaviors of $\sim L^{-6}$ and $\sim L^{-7}$ in the inertial case, exhibit the same behavior of $\sim L^{-4}$, regardless of  whether $L\ll\lambda$ or $L\gg\lambda$, as long as the condition $L\gg\frac{\lambda}{L}L_a$ is satisfied.

In fact, this similar behavior of the interaction potential in the acceleration case has already been observed for two cross-polarizable atoms. The interaction potential for two such atoms displays the common behavior
$-\frac{9\omega^3}{8\pi^2a^2L^6}$ in both the vdW and CP regions $L\ll\lambda$ and $L\gg\lambda$ once $L\gg\frac{\lambda}{L}L_a$~\cite{Zhou25}.

Noteworthily, $L_a\equiv a^{-1}$ is the characteristic length marking the breakdown of the local inertial frame approximation, and the above results indicate that $L_a$ plays a crucial role in determining the scaling behavior of the interatomic interaction potential. For example, by comparing Tabs. I-III, we observe that in vdW subregions I and II (where $L\ll L_a$), the leading-order diagonal components exhibit the same $\sim L^{-6}$ scaling as in the inertial case. In contrast, in subregion III where $L\gg L_a$,  entirely new scalings, such as $\sim a^{-2}L^{-8}$, $\sim L^{-4}$, and $\sim a^{-3}L^{-8}$, emerge. Similarly, in the CP regime, diagonal components in subregion IV where $L\ll L_a$ scale as $\sim L^{-7}$, consistent with the inertial result, whereas those in subregions V and VI where $L\gg L_a$ show new scalings such as $\sim a^{-1}L^{-8}$,  $\sim a^{-2}L^{-8}$, and $\sim L^{-4}$. These observations indicate that inertial-like behavior is retained when $L\ll L_a$, while acceleration-induced effects dominate when $L\gg L_a$.

To summarize, the accelerated motion of atoms introduces significant changes to the interatomic interaction potential compared to the inertial case. However, these effects are distinctive from those of a thermal bath, despite the analogy drawn from the Fulling-Davies-Unruh (FDU) effect, which suggests that a uniformly accelerated atom behaves as if it were in a thermal bath. These distinctions arise in several ways.

First, as analyzed in the section below Eq.~(\ref{potential-general}), off-diagonal components $(\delta E)_q^{jk}$ with $j\neq k$ (or equivalently cross-terms characterized by $|\mu^A_x|^2_{ge}|\mu^B_z|^2_{ge}$ and $|\mu^A_z|^2_{ge}|\mu^B_x|^2_{ge}$)  contribute to the interaction potential. This suggests that an interaction exists between two cross-polarizable atoms in the acceleration case, while such an interaction does not emerge for two cross-polarizable static atoms in a thermal bath~\cite{Cheng23}. The physical origin of this distinction lies in the behavior of the electric field correlation functions. In the acceleration case, correlations of the form  $\langle 0|E_j(x_A(\tau))E_k(x_B(\tau'))|0\rangle$ with $j,k=x,z$ or $z,x$  are nonzero due to the breaking of translational invariance. In contrast, in the thermal case, translational invariance is preserved, and these cross-component correlations vanish. Consequently, off-diagonal terms in the interaction potential arise only in the acceleration case, marking a fundamental difference from thermal effects.

Second, the separation-dependence of the interaction potential in the acceleration case can be dramatically different from that in the thermal case. For example, the potential of two static isotropically polarizable atoms in a thermal bath scales as  $TL^{-6}$ (where $T$ denotes the temperature of the thermal bath)  in the CP region $\lambda\ll T^{-1}\ll L$~\cite{Cheng23}, whereas  in the acceleration case, the potential scales as $T_U^{-1}L^{-8}$  in the corresponding region $\lambda\ll L_a\ll L$ where $L_a=(2\pi T_U)^{-1}=a^{-1}$ (see Eq.~(\ref{V-iso})).

\section{Summary}
We studied the effects of acceleration  on the interatomic interaction by calculating, with the fourth-order DDC formalism, the separate contributions of  vacuum fluctuations of electromagnetic fields (vf-contribution) and  radiation reaction of  atoms (rr-contribution) to the interaction potential between two ground-state atoms. These atoms undergo synchronous and uniform acceleration along  the $x-$direction, maintain a constant separation $L$ along the $z-$direction,  and couple to the fluctuating  electromagnetic vacuum field. The interaction potential is found to be crucially dependent on the atomic polarizability, interatomic separation, and the acceleration. It includes both diagonal components
$(\delta E)^{jk}$ with $j=k$,  which exist in both inertial and acceleration cases,  and off-diagonal components $(\delta E)^{jk}$ with $j\neq k$,  which arise exclusively due to acceleration and vanish in the inertial case. We show that the accelerated motion induces either minor modifications or entirely novel behaviors depending on the region of parameter space.

In the first van der Waals (vdW) subregion where $L\ll\lambda\ll L_a$, the rr-contribution dominates over the vf-contribution. The leading-order diagonal components $(\delta E)^{jk}$ with $j=k$ display the same $L^{-6}$-separation dependence as those in the inertial case, while the off-diagonal components which do not exist in the inertial case behave as $\sim a^2L^{-4}$. In the second vdW subregion $L\ll L_a\ll\lambda$, acceleration-induced modifications to the diagonal components become more pronounced, but the leading-order terms of both diagonal and off-diagonal components remain unchanged.   Consequently, the total interaction potential for isotropically polarizable atoms in these two subregions retains the same $\sim L^{-6}$ dependence as that in the inertial case.

However, in the third vdW subregion $L_a\ll L\ll\lambda$, where acceleration becomes significant, all components of the interaction potential are strongly modified. The interaction potential for isotropically polarizable atoms then exhibits two distinct behaviors depending on the scale: it scales as  $\sim a^{-2}L^{-8}$ in the first subregion of region III (where $L_a\ll L\ll\sqrt{\lambda L_a}$), and as $\sim a^0L^{-4}$ in the second subregion $\sqrt{\lambda L_a}\ll L\ll\lambda$.

In the Casimir-Polder (CP) regime, we considered three subregions.   In the first ($\lambda\ll L\ll L_a$) and the second ($\lambda\ll L_a\ll L$) CP subregions, both the vf- and rr-contributions contain oscillatory and non-oscillatory terms as  interatomic separation varies,  with the oscillatory parts dominating in magnitude. As a result, each of these contributions alone may lead to either attractive or repulsive forces, depending on the separation. However, due to perfect cancellation between the oscillatory terms in the vf- and rr-contributions, the total interaction potential exhibits a smooth, monotonic dependence on separation. In the first CP subregion, the diagonal components of the interaction potential scale as $\sim a^0L^{-7}$, consistent with the inertial case,  while the off-diagonal components which do not show up in the inertial case scale as  $\sim a^2L^{-5}$.   In the second CP subregion, both diagonal and off-diagonal components are acceleration-dependent, scaling respectively as $\sim a^{-1}L^{-8}$ and $\sim a^{-3}L^{-10}$. The total interaction potential of two isotropically polarizable atoms behaves in the leading order as $\sim L^{-7}$  in the first subregion, and as $\sim a^{-1}L^{-8}$ in the second, an entirely new behavior not present in the inertial case.

In the third CP subregion ($L_a\ll\lambda\ll L$), both vf- and rr-contributions become monotonic in separation. Moreover, the rr-contribution dominates, and all components of the interaction potential are strongly modified:   $(\delta E)_{tot}^{xx}$ scales as $\sim a^{-2}L^{-8}$, $(\delta E)_{tot}^{yy}$ and $(\delta E)_{tot}^{zz}$  as $\sim a^0L^{-4}$, and $(\delta E)_{tot}^{xz}$ and $(\delta E)_{tot}^{zx}$  as $\sim a^{-2}L^{-6}$. Interestingly, the total interaction potential of two isotropically polarizable atoms in this region behaves as $\sim-\frac{\omega^3}{16\pi^2L^4}$, which is identical to that in the furthest vdW subregion $\sqrt{\lambda L_a}\ll L\ll\lambda$. This coincidence indicates that, in the acceleration case, both the vdW and CP interaction potentials, despite their distinct behaviors in the inertial case, converge to a common form of $\sim L^{-4}$ under the condition  $L\gg\frac{\lambda}{L}L_a$.

Finally, we emphasize that the acceleration-induced effects on the interatomic interaction are fundamentally different from those induced by a thermal bath. While the Fulling-Davies-Unruh (FDU) effect implies a similarity between uniformly accelerated detectors and static ones in thermal environments, our results show that, at the level of interatomic interactions, acceleration and thermalization yield qualitatively different phenomena. These differences manifest, for example, in the emergence of off-diagonal components in the acceleration case, which are absent in thermal scenarios, and in the contrasting scaling behaviors of the interaction potentials in the CP region.

\begin{acknowledgments}
This work was supported in part by the NSFC under Grants No. 12075084, No. 11690034, No. 11875172, No. 12047551 and No. 12105061, and the K. C. Wong Magna Fund in Ningbo University.
\end{acknowledgments}

\begin{appendix}

\section{Detailed expressions of $f_{jk}(a,L)$ and $g_{jk}(\omega,a,L)$.}\label{coefficient-functions}
The nonzero components of $f_{jk}(a,L)$ and $g_{jk}(\omega,a,L)$ are
\begin{eqnarray}\label{f}
\left\{
  \begin{array}{ll}
    f_{xx}=-\frac{1+a^2L^2}{8\pi^2 L^2\mathcal{N}^2(a,L)}\;,\\
    f_{yy}=-\frac{1+\frac{1}{2}a^2L^2}{8\pi^2L^2\mathcal{N}(a,L)}\;,\\
    f_{zz}=\frac{1+\frac{1}{8}a^2L^2+\frac{1}{16}a^4L^4}{4\pi^2 L^2\mathcal{N}^2(a,L)}\;,\\
    f_{xz}=-f_{zx}=\frac{a\left(1-\frac{1}{2}a^2L^2\right)}{16\pi^2 L\mathcal{N}^2(a,L)}\;,
  \end{array}
\right.
\end{eqnarray}
and
\begin{eqnarray}\label{g}
\quad
\left\{
  \begin{array}{ll}
    g_{xx}=\frac{1+\frac{1}{2}a^2L^2+\frac{1}{4}a^4L^4-\omega^2 L^2\left(1+\frac{1}{4}a^2L^2\right)}{8\pi^2L^3\mathcal{N}^{5/2}(a,L)}\;,\\
    g_{yy}=\frac{1-\omega^2 L^2\left(1+\frac{1}{4}a^2L^2\right)}{8\pi^2L^3\mathcal{N}^{3/2}(a,L)}\;,\\
    g_{zz}=-\frac{1+\frac{5}{8}a^2L^2-\frac{1}{8}\omega^2a^2L^4\left(1+\frac{1}{4}a^2L^2\right)}{4\pi^2L^3\mathcal{N}^{5/2}(a,L)}\;,\\
    g_{xz}=-g_{zx}=-\frac{a(1+a^2L^2+\omega^2L^2\left(1+\frac{1}{4}a^2L^2)\right)}{16\pi^2L^2\mathcal{N}^{5/2}(a,L)}\;,
  \end{array}
\right.
\end{eqnarray}
where we have abbreviated $f_{jk}(a,L)$ and $g_{jk}(\omega,a,L)$ by $f_{jk}$ and $g_{jk}$, respectively.

\section{Detailed expressions of $g_s(x)$ in Tab.~\ref{tab4} and $h_s\left(x,y\right)$ in Tab.~\ref{tab5}.}\label{gh}
The $g_s(x)$ in Tab.~\ref{tab4} with $s$ being a series of integers ranging from $1$ to $9$ are given by
\begin{eqnarray}
\left\{
  \begin{array}{ll}
   g_1(x)=1-x^{-2}+x^{-4}\;,\\
   g_2(x)=1-\frac{4}{9}x^{-2}+\frac{1}{3}x^{-4}\;,\\
   g_3(x)=1-\frac{33}{2}x^{-2}+9x^{-4}+\frac{3}{2}x^{-6}\;,\\
   g_4(x)=1-\frac{10}{3}x^{-2}+3x^{-4}\;,\\
   g_5(x)=1-\frac{15}{2}x^{-2}+\frac{9}{2}x^{-6}\;,\\
   g_6(x)=1-x^{-2}\;,\\
   g_7(x)=1-\frac{1}{3}x^{-2}\;,\\
   g_8(x)=1+\frac{3}{2}x^{-2}\;,\\
   g_9(x)=1+\frac{5}{7}x^{-2}-\frac{1}{7}x^{-4}\;.
\end{array}
\right.
\end{eqnarray}

The $h_s\left(x,y\right)$ in Tab.~\ref{tab5} with $s$ being a series of integers ranging from $1$ to $16$ are given by
\begin{eqnarray}
\left\{\begin{array}{ll}
h_1(x,y)=1-\frac{3}{16\ln{y}}+(\frac{3}{8\ln{y}}-1)x^2+\frac{x^4}{16\ln{y}}\;,\\
h_2(x,y)=1+2(\frac{1}{\ln{y}}-3)x^2+x^4\;,\\
h_3(x,y)=1-\frac{3}{8\ln{y}}+\frac{x^2}{8\ln{y}}\;,\\
h_4(x,y)=1-\frac{5}{8\ln{y}}+\frac{x^2}{8\ln{y}}\;,\\
h_5(x,y)=1-\frac{23}{36\ln{y}}+(\frac{7}{36\ln{y}}-\frac{2}{9})x^2\;,\\
h_6(x,y)=1+(\frac{1}{\ln{y}}-1)x^2\;,\\
h_7(x,y)=1-\frac{1}{4\ln{y}}+(\frac{5}{4\ln{y}}-1)x^2\;,\\
h_8(x,y)=1-\frac{11}{32\ln{y}}+(\frac{47}{32\ln{y}}-\frac{7}{4})x^2
\end{array}
\right.
\end{eqnarray}
and
\begin{eqnarray}
\left\{\begin{array}{ll}
h_9(x,y)=1-\frac{11}{24\ln{y}}+\frac{x^2}{8\ln{y}}\;,\\
h_{10}(x,y)=1-\frac{33}{76\ln{y}}+(\frac{43}{228\ln{y}}-\frac{10}{57})x^2\;,\\
h_{11}(x,y)=1-\frac{1}{12\ln{y}}+(\frac{13}{12\ln{y}}-1)x^2\;,\\
h_{12}(x,y)=1-\frac{9}{64\ln{y}}+(\frac{85}{64\ln{y}}-\frac{43}{24})x^2\;,\\
h_{13}(x,y)=1-\frac{11}{32\ln{y}}+(\frac{7}{32\ln{y}}-\frac{1}{4})x^2\;,\\
h_{14}(x,y)=1-\frac{107}{308\ln{y}}+(\frac{95}{308\ln{y}}-\frac{38}{17})x^2+\frac{x^4}{77\ln{y}}\;,\\
h_{15}(x,y)=1-\frac{1}{12\ln{y}}+(\frac{17}{12\ln{y}}-\frac{7}{3})x^2\;,\\
h_{16}(x,y)=1-\frac{9}{64\ln{y}}+(\frac{367}{192\ln{y}}-\frac{29}{8})x^2+\frac{x^4}{6}\;.
\end{array}
\right.
\end{eqnarray}
\end{appendix}


\begin{thebibliography}{}
\bibitem{Craig98} D. P. Craig and T. Thirunamachandran, {\it Molecular Quantum Electrodynamics} (Dover, Mineola, NY, 1998).
\bibitem{Israelachvili73} J. N. Israelachvili, {\it Van der Waals forces in biological systems}, \href{https://doi.org/10.1017/S0033583500001566}{Q. Rev. Biophys. {\bf 6}, 341 (1973)}.
\bibitem{Roth96} C. M. Roth, B. L. Neal, and A. M. Lenhoff, {\it Van der Waals interactions involving proteins}, \href{https://doi.org/10.1016/S0006-3495(96)79641-8}{Biophys. J {\bf 70}, 977 (1996)}.
\bibitem{Bloch08} I. Bloch, J. Dalibard, and W. Zwerger, {\it Many-body physics with ultra cold gases}, \href{https://doi.org/10.1103/RevModPhys.80.885}{Rev. Mod. Phys. {\bf 80}, 885 (2008)}.
\bibitem{Tkatchenko11} A. Tkatchenko, M. Rossi, V. Blum, J. Ireta, and M. Scheffler, {\it Unraveling the stability of polypeptide helices: critical role of van der Waals interactions}, \href{https://doi.org/10.1103/PhysRevLett.106.118102}{Phys. Rev. Lett. {\bf 106}, 118102 (2011)}.
\bibitem{Woods16} L. M. Woods, D. A. R. Dalvit, A. Tkatchenko, P. Rodriguez-Lopez, A. W. Rodriguez, and R. Podgornik, {\it Materials perspective on Casimir and van der Waals interactions}, \href{https://doi.org/10.1103/RevModPhys.88.045003}{Rev. Mod. Phys. {\bf 88}, 045003 (2016)}.
\bibitem{Defenu23} N. Defenu, T. Donner, T. Macr$\grave{\i}$, G. Pagano, S. Ruffo, and A. Trombettoni, {\it Long-range interacting quantum systems}, \href{https://doi.org/10.1103/RevModPhys.95.035002}{Rev. Mod. Phys. {\bf 95}, 035002 (2023)}.
\bibitem{Klimchitskaya09} G. L. Klimchitskaya, U. Mohideen, and V. M. Mostepanenko, {\it The Casimir force between real materials: experiment and theory}, \href{https://doi.org/10.1103/RevModPhys.81.1827} {Rev. Mod. Phys. {\bf 81}, 1827 (2009)}.
\bibitem{Casimir-Polder48} H. B. G. Casimir and D. Polder, {\it The influence of retardation on the London-van der Waals forces}, \href{https://doi.org/10.1103/PhysRev.73.360}{{Phys. Rev.} {\bf 73}, 360 (1948)}.
\bibitem{Power82} E. A. Power and T. Thirunamachandran, {\it Quantum electrodynamics in a cavity}, \href{https://journals.aps.org/pra/abstract/10.1103/PhysRevA.25.2473}{Phys. Rev. A {\bf 25}, 2473 (1982)}.
\bibitem{Spagnolo06} S. Spagnolo, R. Passante, and L. Rizzuto, {\it Field fluctuations near a conducting plate and Casimir-Polder forces in the presence of boundary conditions}, \href{https://journals.aps.org/pra/abstract/10.1103/PhysRevA.73.062117}{Phys. Rev. A {\bf 73}, 062117 (2006)}.
\bibitem{Passante07} R. Passante and S. Spagnolo, {\it Casimir-Polder interatomic potential between two atoms at finite temperature and in the presence of boundary conditions}, \href{https://journals.aps.org/pra/abstract/10.1103/PhysRevA.76.042112}{Phys. Rev. A {\bf 76}, 042112 (2007)}.
\bibitem{Peng23} Y. Peng, S. Cheng, and W. Zhou, {\it Interaction potential of two nonidentical ground-state atoms}, \href{https://doi.org/10.1088/1572-9494/acd4a5}{Commun. Theor. Phys. {\bf 75}, 085102 (2023)}.
\bibitem{McLachlan63} A. D. McLachlan, {\it Retarded dispersion forces in dielectrics at finite temperatures}, \href{https://royalsocietypublishing.org/doi/abs/10.1098/rspa.1963.0115}{Proc. R. Soc. Lond. A {\bf 274}, 80 (1963)}.
\bibitem{Boyer96} T. H. Boyer, {\it Temperature dependence of Van der Waals forces in classical electrodynamics with classical electromagnetic zero-point radiation}, \href{https://journals.aps.org/pra/abstract/10.1103/PhysRevA.11.1650}{Phys. Rev. A {\bf 11}, 1650 (1975).}
\bibitem{Milonni96} P. W. Milonni and A. Smith, {\it Van der Waals dispersion forces in electromagnetic fields}, \href{https://journals.aps.org/pra/abstract/10.1103/PhysRevA.53.3484}{Phys. Rev. A {\bf 53}, 3484 (1996)}.
\bibitem{Ninham98} B. W. Ninham and J. Daicic, {\it Lifshitz theory of Casimir forces at finite temperature}, \href{https://journals.aps.org/pra/abstract/10.1103/PhysRevA.57.1870}{Phys. Rev. A {\bf 57}, 1870 (1998)}.
\bibitem{Wennerstrom99} H. Wennerstr$\ddot{o}$m, J. Daicic, and B. W. Ninham, {\it Temperature dependence of atom-atom interactions}, \href{https://journals.aps.org/pra/abstract/10.1103/PhysRevA.60.2581}{Phys. Rev. A {\bf 60}, 2581 (1999)}.
\bibitem{Goedecke99} G. H. Goedecke and R. C. Wood, {\it Casimir-Polder interaction at finite temperature}, \href{https://journals.aps.org/pra/abstract/10.1103/PhysRevA.60.2577}{Phys. Rev. A {\bf 60}, 2577 (1999)}.
\bibitem{Cheng221} S. Cheng, W. Zhou, and H. Yu, {\it Quantum thermal field fluctuation induced corrections to the interaction between two ground-state atoms}, \href{https://doi.org/10.1088/1572-9494/ac8f2d}{Commun. Theor. Phys. {\bf 74}, 125103 (2022)}.
\bibitem{Cheng23} S. Cheng, W. Zhou, and H. Yu, {\it General framework for interatomic interaction energy of two ground-state atoms in a thermal bath}, \href{https://journals.aps.org/pra/abstract/10.1103/PhysRevA.107.012815}{Phys. Rev. A {\bf 107}, 012815 (2023)}.
\bibitem{Fulling73} S. A. Fulling, {\it Nonuniqueness of Canonical Field Quantization in Riemannian Space-Time}, \href{https://doi.org/10.1103/PhysRevD.7.2850}{Phys. Rev. D {\bf 7}, 2850 (1973)}.
\bibitem{Davies75} P. C. W. Davies, {\it Scalar production in Schwarzschild and Rindler metrics}, \href{https://iopscience.iop.org/article/10.1088/0305-4470/8/4/022/meta?casa_token=zToqZxRw2_gAAAAA:ooJvibEQCTeBY7e1s5OuNjQ1vijrWy40-TbCE5GrrEKjaRXjaIVMIg2GBmz8W3N1CBJQhJuhVHI6K8NbVA}{J. Phys. A {\bf 8}, 609 (1975)}.
\bibitem{Unruh76} W. G. Unruh, {\it Notes on black-hole evaporation}, \href{https://doi.org/10.1103/PhysRevD.14.870} {Phys. Rev. D {\bf 14}, 870 (1976)}.
\bibitem{Marino14} J. Marino, A. Noto, and R. Passante, {\it Thermal and nonthermal signatures of the Unruh effect in Casimir-Polder forces}, \href{https://doi.org/10.1103/PhysRevLett.113.020403}{Phys. Rev. Lett. {\bf 113}, 020403 (2014)}.
\bibitem{Cheng22} S. Cheng, W. Zhou, and H. Yu, {\it Probing long-range properties of vacuum altered by uniformly accelerating two spatially separated Unruh-DeWitt detectors}, \href{https://doi.org/10.1103/PhysRevLett.113.020403}{Phys. Lett. B {\bf 834}, 137440 (2022)}.
\bibitem{Zhou25} W. Zhou, S. Cheng, and H. Yu, {\it Interaction between Unruh-Dewitt detectors exclusively due to acceleration: A parallel to the FDU effect}, \href{https://doi.org/10.1016/j.physletb.2025.139271}{Phys. Lett. B {\bf 862}, 139271 (2025)}.
\bibitem{Greiner} W. Greiner and J. Reinhardt, {\it Field Quantization} (Springer Verlag, Berlin, 1996), p. 177.
\bibitem{DDC82} J. Dalibard, J. Dupont-Roc, and C. Cohen-Tannoudji, {\it Vacuum fluctuations and radiation reaction: identification of their respective contributions}, \href{https://doi.org/10.1051/jphys:0198200430110161700}
                {J. Phys. France {\bf 43}, 1617 (1982)}.
\bibitem{DDC84} J. Dalibard, J. Dupont-Roc, and C. Cohen-Tannoudji, {\it Dynamics of a small system coupled to a reservoir: reservoir fluctuations and self-reaction}, \href{https://doi.org/10.1051/jphys:01984004504063700}
                {J. Phys. France {\bf 45}, 6376 (1984)}.
\bibitem{Zhou21} W. Zhou, S. Cheng, and H. Yu, {\it Interatomic interaction of two ground-state atoms in vacuum: contributions of vacuum fluctuations and radiation reaction}, \href{https://doi.org10.1103/PhysRevA.103.012227}{Phys. Rev. A {\bf 103}, 012227 (2021)}.
\bibitem{Meschede90} D. Meschede, W. Jhe, and E. A. Hinds, {\it Radiative properties of atoms near a conducting plane: An old problem in a new light}, \href{https://doi.org/10.1103/PhysRevA.41.1587}{{Phys. Rev. A} {\bf 41}, 1587 (1990)}.
\bibitem{Audretsch94} J. Audretsch and R. M$\ddot{\text{u}}$ller, {\it Spontaneous excitation of an accelerated atom: The contributions of vacuum fluctuations and radiation reaction}, \href{https://doi.org/10.1103/PhysRevA.50.1755}{Phys. Rev. A {\bf 50}, 1755 (1994)}.
\bibitem{Holzmann95} J. Audretsch, R. M$\ddot{\text{u}}$ller, and M. Holzmann, {\it Generalized Unruh effect and Lamb shift for atoms on arbitrary stationary trajectories}, \href{https://iopscience.iop.org/article/10.1088/0264-9381/12/12/010/meta}{Class. Quant. Grav. {\bf 12}, 2927 (1995)}.
\bibitem{Tomazelli03} J. L. Tomazelli and L. C. Costa, {\it Atomic radiative transitions in thermo field dynamics}, \href{https://doi.org/10.1142/S0217751X03012345}{Int. J. Mod. Phys. A {\bf 18}, 1079 (2003)}.
\bibitem{Yu05} H. Yu and S. Lu, {\it Spontaneous excitation of an accelerated atom in a spacetime with a reflecting plane boundary}, \href{https://doi.org/10.1103/PhysRevD.72.064022}{Phys. Rev. D {\bf 72}, 064022 (2005)}; \href{https://doi.org/10.1103/PhysRevD.73.109901}{{\bf 73}, 109901(E) (2006)}.
\bibitem{Zhu06} Z. Zhu, H. Yu, and S. Lu, {\it Spontaneous excitation of an accelerated hydrogen atom coupled with electromagnetic vacuum fluctuations}, \href{https://doi.org/10.1103/PhysRevD.73.107501} {Phys. Rev. D {\bf 73}, 107501 (2006)}.
\bibitem{Yu06} H. Yu and Z. Zhu, {\it Spontaneous absorption of an accelerated hydrogen atom near a conducting plane in vacuum}, \href{https://doi.org/10.1103/PhysRevD.74.044032}{Phys. Rev. D {\bf 74}, 044032 (2006)}.
\bibitem{Zhou12} W. Zhou and H. Yu, {\it Spontaneous excitation of a uniformly accelerated atom coupled to vacuum Dirac field fluctuations}, \href{https://doi.org/10.1103/PhysRevA.86.033841}{Phys. Rev. A {\bf 86}, 033841 (2012)}.
\bibitem{Audretsch95} J. Audretsch and R. M$\ddot{\text{u}}$ller, {\it Radiative energy shifts of an accelerated two-level system}, \href{https://doi.org/10.1103/PhysRevA.52.629}{Phys. Rev. A {\bf 52}, 629 (1995)}.
\bibitem{Passante98} R. Passante, {\it Radiative level shifts of an accelerated hydrogen atom and the Unruh effect in quantum electrodynamics}, \href{https://doi.org/10.1103/PhysRevA.57.1590}{Phys. Rev. A {\bf 57}, 1590 (1998)}.
\bibitem{Rizzuto07} L. Rizzuto, {\it Casimir-Polder interaction between an accelerated two-level system and an infinite plate}, \href{https://doi.org/10.1103/PhysRevA.76.062114}{Phys. Rev. A {\bf 76}, 062114 (2007)}.
\bibitem{Rizzuto09} L. Rizzuto and S. Spagnolo, {\it Lamb shift of a uniformly accelerated hydrogen atom in the presence of a conducting plate}, \href{https://doi.org/10.1103/PhysRevA.79.062110}{Phys. Rev. A {\bf 79}, 062110 (2009)}.
\bibitem{Zhu10} Z. Zhu and H. Yu, {\it Position-dependent energy-level shifts of an accelerated atom in the presence of a boundary}, \href{https://doi.org/10.1103/PhysRevA.82.042108}{Phys. Rev. A {\bf 82}, 042108 (2010)}.
\bibitem{Zhou16} W. Zhou, R. Passante, and L. Rizzuto, {\it Resonance interaction energy between two accelerated identical atoms in a coaccelerated frame and the Unruh effect}, \href{https://doi.org/10.1103/PhysRevD.94.105025}{Phys. Rev. D {\bf 94}, 105025 (2016)}.
\bibitem{Rizzuto16} L. Rizzuto, M. Lattuca, J. Marino, A. Noto, S. Spagnolo, W. Zhou, and R. Passante, {\it Nonthermal effects of acceleration in the resonance interaction between two uniformly accelerated atoms}, \href{https://doi.org/10.1103/PhysRevA.94.012121}{Phys. Rev. A {\bf 94}, 012121 (2016)}.
\bibitem{London30} F. London, Z. Phys. {\bf 63}, 245 (1930); Z. Phys. Chem. Abt. B {\bf 11}, 222 (1930).
\end{thebibliography}
\end{document}